\documentstyle[amsfonts,prl,aps]{revtex}

\input{psfig.sty}

\begin{document}
\draft

\twocolumn[
\hsize\textwidth\columnwidth\hsize\csname @twocolumnfalse\endcsname
\title{Electron - hole asymmetry and activation energy in quantum Hall ferromagnets}
\author{Vladimir I. Fal'ko $^\ast$, S.V. Iordanski $^{\ddagger \ast }$}
\address{$^{\ast }$ Physics Department, Lancaster University, LA1
4YB, UK\\
$^{\ddagger }$ Landau Institute for Theoretical Physics, ul. Kosygina 1, 
Moscow, Russia}
\date{today}

\maketitle

\begin{abstract}
We argue that the dissipative transport in ferromagnetic quantum  
Hall effect liquids at $\nu=2N+1$ is dominated by the thermal activation
of pairs consisting of an electron and an antiskyrmion (topological texture 
which represents a hole with 'screened' exchange interaction),
thus manifesting the lack of electron-hole symmetry in 
quantum Hall ferromagnets. We find that the activation energy 
of such a pair is not the exchange energy, but is determined by
the interplay between the excess Zeeman energy of a skyrmion 
and the charging energy of its topological texture. 

\end{abstract}
\narrowtext]

Due to the electron-electron exchange interaction, the system of
two-dimensional (2D) electrons filling completely an odd-integer number of
spin-polarized Landau levels (LL) form an incompressible liquid with
ferromagnetic properties \cite{Girvin,Lee}. The incompressibility of such a
liquid is usually studied experimentally by measuring the activation energy, 
${\cal E}$, in the dissipative part of the conductivity tensor, $\sigma
_{xx}\propto e^{-{\cal E}/2T}$. The activation energy (at, for instance, $%
\nu =1$) reflects the enhancement of a gap in the excitation spectrum of
quantum Hall ferromagnets (QHF's), as compared to the bare Zeeman energy of
2D electrons, $\varepsilon _{{\rm Z}}$. For a long time, it was believed 
\cite{Ando}\ that the experimentally observed ${\cal E}$ is nothing but the
activation energy, $\Im _{0}$, of a pair consisting of a free electron and a
hole at the filled spin-polarized Landau level, which is determined by the
intra-LL exchange interaction ($\Im _{0}=\int dzV(\lambda \sqrt{2z}%
)e^{-z}\approx \sqrt{\pi /2}(e^{2}/\lambda \chi )$ \ for $\nu =1$, $\lambda =%
\sqrt{\hbar c/eB}$).

However, there is growing experimental evidence that the energy required for
activation of a dissipative current in QHF's in 2D systems with a reduced
value of bare Zeeman splitting is much less than $\Im _{0}$. The
accumulation of this evidence is accompanied by extensive theoretical
studies of skyrmions \cite{Sondhi,MacDonald}\ - charge carriers whose
exchange interaction with the ground state electrons is 'screened' by a
topologically stable spin polarization texture. For $\varepsilon _{{\rm Z}%
}\ll \Im _{0}$, the activation energy of a skyrmion - antiskyrmion pair is
determined by a smaller inter-LL exchange energy ($\Im _{i}=\int dzV(\lambda 
\sqrt{2z})e^{-z}{\rm L}_{1}(z)\approx \frac{1}{2}\sqrt{\pi /2}(e^{2}/\lambda
\chi )$ for $\nu =1$), which is the result of a local twist of the
spin-polarization field and requires the mixing of different Landau levels.

Nevertheless, even despite the factor of two reduction, the activation
energy of a skyrmion-antiskirmion pair is too large to explain the
experimentally observed tendency \cite
{Maude,Leadley,Leadley1,Shayegan,Dolgopolov} of a gap in quantum Hall
ferromagnets with a small Zeeman energy to diminish, when $\varepsilon _{%
{\rm Z}}\rightarrow 0$. In the present paper, we propose an alternative
scenario for the activation of charge carriers in QHF's by exploiting the
lack of electron-hole symmetry in such a system. Below, we argue that a
dissipative current in QHF's (in the thermodynamic limit) is provided by
thermal activation of a couple consisting of one single-particle electron
and an antiskyrmion (the latter represents a hole with an exchange
interaction 'screened' by a topological texture). Such an asymmetric pair
has activation energy 
\begin{equation}
{\cal E}=a\varepsilon _{{\rm Z}}^{1/3}E_{{\rm C}}^{2/3}\ln ^{1/3}\left( 
\frac{\Im _{i}}{E_{{\rm C}}^{2/3}\varepsilon _{{\rm Z}}^{1/3}}\right)
,\;\;E_{{\rm C}}=\frac{e^{2}}{\chi \lambda },  \label{Result}
\end{equation}
(with$\;a\approx 0.9$), which is much smaller than even the inter-LL
exchange energy. The same quantity, ${\cal E}$ also determines the jump in
the chemical potential of the 2D system across the unit filling factor.
Together with the idea of the electron-'hole' asymmetry, Eq. (\ref{Result})
for the transport activation energy represents the main result of this paper.

First, we would like to comment on the idea of the electron-hole asymmetry
of thermodynamic properties of the QHF. When analyzing QHF's, it is very
tempting to restrict the microscopic analysis to only a couple of spin-split
Landau levels \cite{MacDonald} (spin-up/down $n=0$ for $\nu =1$). After such
a premature projection onto the lowest LL, the symmetry between a
single-particle electron at the excited-spin LL and a hole at a completely
filled one is artificially created, and it dominates in the further analysis
making all thermodynamic properties of a liquid symmetric {\it ad hoc} with
respect to the electron-hole transformation. In particular, this places the
energies of an electron/hole and also of a pair of a skyrmion and
antiskyrnion symmetric with respect to the chemical potential of a liquid,
which has, then, to be placed half-way between the energies of equivalent
spin-split carriers. However, a more cautious analysis of the skyrmionic
texture energy requires the mixing of Landau levels by the gradients of the
exchange field and cannot be obtained without consistently taking into
account other LL's. Taking account of the higher LL's can be done following
various roots, and it results in both spin rigidity of a polarized liquid
and in a topological term in its energy which strongly distinguishes between
skyrmionic excitations corresponding to the addition or subtraction of a
physical electron.

The formation of a skyrmion aims to reduce the exchange energy of a charged
carrier (an electron or a hole at the filled LL) by smoothly adjusting the
spinor wave functions of neighboring electrons in such a way that in the
locally rotated spin-coordinate system there would be one completely filled
spin-polarized LL. We describet he transformation $\left( 
\begin{array}{c}
\varphi ({\bf r}) \\ 
0
\end{array}
\right) =U({\bf r})\psi ({\bf r})$ to the local spin-coordinate system,
where the spin-density matrix is diagonal $\Lambda =(1+\sigma ^{3})/2$,
using unitary rotations, $U$ from the group U(2) \cite{RemU2}. After such a
tranformation, the LL in the rotated spin-coordinate system have to be
slightly modified by taking into account an additional non-Abelian gauge
field, 
\begin{equation}
\vec{\Omega}({\bf r})=U({\bf r})\left( -i\vec{\partial}U^{\dagger }({\bf r}%
)\right) \equiv \left( i\vec{\partial}U({\bf r})\right) U^{\dagger }({\bf r}%
),  \label{21}
\end{equation}
which has all three independent spin-matrix components: $\vec{\Omega}({\bf r}%
)=\vec{a}^{\alpha }\sigma ^{\alpha }$ ($\sigma ^{1,2,3}$ stand for Pauli
matrices). One of those three, defined as $\vec{a}^{3}({\bf r})={\rm tr}%
(\Lambda \vec{\Omega}({\bf r}))$, produces the same effect, as an additional
'magnetic field' \cite{RemU2}, 
\begin{equation}
b=\frac{c}{e}\left[ \partial _{x}a_{y}^{3}-\partial _{y}a_{x}^{3}\right] =i%
\frac{c}{e}{\rm tr}\left( \Lambda \left[ \Omega _{x},\Omega _{y}\right]
\right) ,  \label{density1}
\end{equation}
which changes the capacity to accommodate electrons in one
'gauge-transformed' LL (as compared to a homogeneously polarized one) by $%
q=\int \rho ({\bf r})d{\bf r}$ real particles, where \ 
\begin{equation}
\rho =\frac{\theta i}{2\pi }{\rm tr}\left( \Lambda \left[ \Omega _{x},\Omega
_{y}\right] \right) ,\;{\rm and}\;\theta =\frac{eB_{z}}{\left| eB_{z}\right| 
}.
\end{equation}
The latter relation takes into account whether the additional field $b$
should be added, or subtracted from the external magnetic field value, which
is the the first indication the 'electron'-'hole' symmetry in the system is
violated. Note that, for the reversed-spin LL in the rotated frame, the same 
$\vec{\Omega}(r)$ of the same texture in $U({\bf r})$ produces the opposite
effect, so that the total number of states available at one LL remains the
same (${\rm tr}\vec{\Omega}=0$).

The texture with $q=1$ corresponds to one added particle and represents an
electron with 'screened' exchange interaction (skyrmion); $q=-1$ is
analogous to a 'hole' (antiskyrmion). Both of them are topologically stable
textures in the local spin-polarization field of a liquid, $n^{\alpha }={\rm %
tr}(U\sigma ^{\alpha }U^{\dagger }\Lambda )$, which can be classified using
the homotopy group $\pi _{2}({\rm U(2)}/[{\rm U(1)}\times {\rm U(1)}])$,
that is, using the index of mapping of a 2D plane onto a unit sphere, 
\[
\int \frac{d{\bf r}}{8\pi }\epsilon ^{\beta \gamma \delta }\epsilon
_{zij}n^{\beta }\partial _{i}n^{\gamma }\partial _{j}n^{\delta }=-\theta q. 
\]

The presence of a spurious gauge field $\vec{a}^{\alpha }$ also affects the
energy of a liquid, due to the exchange interaction, and it results in two
contributions to the energy of a twisted texture. One is generated by
off-diagonal components $\vec{a}^{1}$ and $\vec{a}^{2}$ of the gauge field $%
\vec{\Omega}$. Since, in the rotated coordinate system, the Landau level is
full and the liquid is polarized along the axis ${\bf l}_{3}$, these two
components enter only as perturbations in the second order, ${\rm tr}\left(
\Lambda \vec{\Omega}(1-\Lambda )\vec{\Omega}\right) $, thus producing the
spin-rigidity term in the energy. This term always increases the energy of a
liquid when the texture is deformed. The other contribution comes from the
diagonal (in the ${\bf l}_{3}$-basis) component of the non-Abelian gauge,
and, as a diagonal term, it depends on the sign of the 'magnetic field' $b$.
In fact, what matters is whether the exchange field of a texture tends to
twirl electrons in the same direction as the external field $B_{z}$, or
opposite to it. Since the relative signs of $b$ and $B_{z}$ determine
whether the texture carries extra electrons, this generates the
'electron'-'hole' asymmetry for excitations in QHF's and makes an additional
difference between the energy shift of a polarized liquid upon adding or
subtracting a real particle (electron).

In the following, we focus our attention on a single-carrier texture and
study its energy using the standard root \cite{Fradkin} of the
Hubbard-Stratonovich transformation applied to the partition function, $%
{\cal Z}=e^{-\Phi /T}$ of the interacting electron gas, 
\begin{eqnarray*}
{\cal Z} &=&\int {\frak D}\psi {\frak D}\psi ^{\dagger }\exp \left\{ -\frac{%
\Phi _{0}}{T}-\int_{0}^{\beta }d\tau E_{{\rm int}}\right\} \\
\frac{\Phi _{0}}{T} &=&\int_{0}^{\beta }d\tau \int d{\bf r}\psi ^{\dagger
}\left( \partial _{\tau }-\mu +H_{0}-\frac{\varepsilon _{{\rm Z}}}{2}\sigma
^{z}\right) \psi \\
H_{0} &=&\frac{1}{2m}(-i\vec{\partial}-\frac{e}{c}\vec{A})^{2},\;\;\beta
=T^{-1},\;\hbar =1, \\
E_{{\rm int}} &=&\frac{1}{2}\int d{\bf r}d{\bf r}^{\prime }V({\bf r}-{\bf r}%
^{\prime })\psi _{\alpha }^{\dagger }({\bf r})\psi _{\alpha }({\bf r})\psi
_{\beta }^{\dagger }({\bf r}^{\prime })\psi _{\beta }({\bf r}^{\prime }),
\end{eqnarray*}
where $\mu $\ is\ chemical potential placed somewhere in the gap of the
single-particle excitation spectrum of the system, fields $\psi $ obey
anti-periodic boundary conditions, as a function of imaginary time $\tau $,
and $V>0$. Then, we expand the thermodynamic potential, $\Phi =\left\langle
E-\mu N\right\rangle $ with respect to the non-Abelian gauge field $\Omega
_{i}({\bf r})$ and its derivatives. The whole procedure is equivalent to the
self-consistent Hartree-Fock calculation, and the only, but essential
difference of this work from what has been published earlier is that (i) we
do not assume a premature projection of electron states onto the lowest
Landau level, and (ii) after we find the presence of a topological term in
the energy, we take it seriously.

The Hubbard-Stratonovich transformation reduces functional integration over
fermionic fields to the functional integration over the exchange mean field $%
Q(\tau ;{\bf r},{\bf r}^{\prime })$, $Q^{\dagger }({\bf r},{\bf r}^{\prime
})=Q({\bf r}^{\prime },{\bf r})$: 
\[
e^{-\int d\tau H_{{\rm int}}}=\int DQe^{-\int d\tau \int d{\bf r}d{\bf r}%
^{\prime }F_{{\rm HS}}} 
\]
where 
\begin{eqnarray*}
F_{{\rm HS}} &=&\frac{\Im _{0}}{2}{\rm tr}Q(\tau ;{\bf r},{\bf r}^{\prime
})Q^{\dagger }(\tau ;{\bf r},{\bf r}^{\prime }) \\
&&+\sqrt{\Im _{0}V({\bf r}-{\bf r}^{\prime })}\psi _{\alpha }^{\dagger }(%
{\bf r})Q_{\alpha \beta }(\tau ;{\bf r},{\bf r}^{\prime })\psi _{\beta }(%
{\bf r}^{\prime }).
\end{eqnarray*}
After integrating out fermions, the thermodynamic potential $\Phi $ can be
represented in the form of 
\begin{eqnarray*}
\frac{\Phi }{T} &=&\int_{0}^{\beta }d\tau \int d{\bf r}d{\bf r}^{\prime }%
\frac{\Im _{0}}{2}Q(\tau ;{\bf r},{\bf r}^{\prime })Q^{\dagger }(\tau ;{\bf r%
},{\bf r}^{\prime }) \\
&&-{\rm Tr}\ln \left\{ H_{0}+\sqrt{\Im _{0}V({\bf r}-{\bf r}^{\prime })}%
Q(\tau ;{\bf r},{\bf r}^{\prime })\right\} ,
\end{eqnarray*}
where Tr stands for the complete trace, including spatial and imaginary time
integrations of a properly ordered product of matrices \cite{Fradkin}. For
comparison, in a homogeneously polarized liquid, without taking into account
the possibility to form topological textures, thermodynamic potential is
equal to 
\begin{equation}
\Phi =|\varepsilon _{Z}|N_{{\rm e}}+\Im _{0}N_{{\rm h}}-\mu \left[ N_{{\rm e}%
}-N_{{\rm h}}\right] .  \label{Eeh}
\end{equation}

When studying the textured state, we use the saddle-point method to find the
value of a matrix $\tilde{Q}=U({\bf r})Q(\tau ;{\bf r},{\bf r}^{\prime
})U^{\dagger }({\bf r}^{\prime })$, which describes the exchange within one
completely filled LL in the rotated spin-frame. The use of a saddle-point $%
\tilde{Q}$ enables us to calculate the exchange part of the thermodynamic
potential of the sea of electrons, 
\[
\frac{\left\langle \Phi \right\rangle _{{\rm ex}}}{T}=\int_{0}^{\beta }d\tau
\int d{\bf r}d{\bf r}^{\prime }\frac{\Im _{0}}{2}\tilde{Q}(\tau ;{\bf r},%
{\bf r}^{\prime })\tilde{Q}^{\dagger }(\tau ;{\bf r},{\bf r}^{\prime
})\;\;\;\;\;\;\;\;\;\; 
\]
\[
-{\rm Tr}\ln \left\{ \partial _{\tau }-\mu +H_{0}+H_{\Omega }+\sqrt{\Im
_{0}V({\bf r}-{\bf r}^{\prime })}\tilde{Q}(\tau ;{\bf r},{\bf r}^{\prime
})\right\} 
\]
for a given texture with a given number of added or missing particles, $q$.
The non-Abelian gauge field $\Omega $ determined by the texture is included
into the perturbation term, 
\[
H_{\Omega }=U({\bf r})H_{0}U^{\dagger }({\bf r})-H_{0}=U({\bf r})\left[
H_{0},U^{\dagger }({\bf r})\right] . 
\]
The saddle point equation for the exchange field of a full LL in the rotated
spin-coordinate system can be obtained after applying the variational
principle to $\left\langle \Phi \right\rangle _{{\rm ex}}$,

\[
\tilde{Q}(\tau ;{\bf r},{\bf r}^{\prime })=\sqrt{\frac{V({\bf r}-{\bf r}%
^{\prime })}{\Im _{0}}}G\left( \tau ,{\bf x};\tau +0,{\bf x}^{\prime
}\right) , 
\]
and, then, solved perturbatively by expanding the Green function, $G\equiv %
\left[ \partial _{\tau }-\mu +H_{0}+H_{\Omega }+\sqrt{\Im _{0}V({\bf r}-{\bf %
r}^{\prime })}\tilde{Q}\right] ^{-1}$ over$\ H_{\Omega }$. An intermediate
step of such a calculation, which is completely equivalent to the
Hartree-Fock procedure, can be schematically represented as

\[
\frac{\left\langle \Phi \right\rangle _{{\rm ex}}}{T}=-{\rm Tr}\left(
H_{\Omega }G_{0}\right) +\frac{1}{2}{\rm Tr}\left( H_{\Omega }G_{0}H_{\Omega
}G_{0}\right) 
\]
\[
-\frac{1}{2}\int d\tau d{\bf r}d{\bf r}^{\prime }{\rm tr}\left\{ \left(
G_{0}H_{\Omega }G_{0}\right) _{\tau {\bf r},\tau {\bf r}^{\prime }}V({\bf r}-%
{\bf r}^{\prime })\left( G_{0}H_{\Omega }G_{0}\right) _{\tau {\bf r}^{\prime
},\tau {\bf r}}\right\} , 
\]
where $G_{0}$ is the electron Green function in a system with one completely
filled homogeneously polarized LL \cite{Rem1} and indices after brackets
indicate the 'outer' coordinates/time, whereas inside each product the full
integration is assumed. For the sake of convenience, we shall use the basis
of LL's and inter-Landau level operators $a_{\mp }=(p_{x}\pm i\theta
p_{y})\lambda /\sqrt{2}$ (where $\vec{p}=-i\vec{\partial}-e\vec{A}{\bf /}c$,
and\ $\theta =eB_{z}/\left| eB_{z}\right| $ indicates the direction of
cyclotron rotation of a carrier), so that

\[
H_{\Omega }=\omega _{{\rm c}}\left\{ a_{+}\Omega _{-}+\Omega
_{+}a_{-}+\Omega _{+}\Omega _{-}\right\} , 
\]
where $\Omega _{\mp }=U({\bf r})\left[ a_{\mp },U^{\dagger }({\bf r})\right]
=(\Omega _{x}\pm i\theta \Omega _{y})\lambda /\sqrt{2}$, and $\omega _{{\rm c%
}}=|eB_{z}|/mc$. \ 

Using integral properties of Hermitian polynomials and recursion relations
for Laguerre polynomials, ${\rm L}_{N}^{m}$, one finally arrives (for $%
\omega _{{\rm c}}\gg \Im $) at

\[
\left\langle \Phi \right\rangle _{{\rm ex}}=\frac{\Im _{i}}{2\pi \lambda ^{2}%
}\int d{\bf r}{\rm tr}\left( \Lambda \Omega _{+}\Omega _{-}-\Lambda \Omega
_{+}\Lambda \Omega _{-}\right) \;\;\;\;\;\;\; 
\]
\[
=\frac{\Im _{i}}{2}\int \frac{d{\bf r}}{2\pi }\left\{ {\rm tr}\left( \Lambda 
\vec{\Omega}(1-\Lambda )\vec{\Omega}\right) +\theta i{\rm tr}\left( \Lambda %
\left[ \Omega _{x},\Omega _{y}\right] \right) \right\} 
\]
where $\Im _{i}$ is the inter-LL exchange, which can be also calculated for
larger odd-integer filling factors, $\nu =2K+1$, as

\begin{eqnarray*}
\Im _{i} &=&\int_{0}^{\infty }dzV(\lambda \sqrt{2z})e^{-z}\times \\
&&\times \left\{ \sum_{M=K,K+1}M\left[ {\rm L}_{M}(z){\rm L}_{K}^{1}(z)-{\rm %
L}_{M-1}(z){\rm L}_{K-1}^{1}(z)\right] \right\}
\end{eqnarray*}

For the sake of a single-charged texture analysis, we use the
parametrization $U=\left( 1+|w|^{2}\right) ^{-1/2}\left( 
\begin{array}{cc}
w & 1 \\ 
-1 & w^{\ast }
\end{array}
\right) $ of a unitary matrix ($w$ can be related to the polarization vector
as $w=(n^{x}+in^{y})/(1-n^{z})$), and $\left\langle \Phi \right\rangle _{%
{\rm ex}}$ takes the form of a conformally-invariant Belavin-Polyakov (BP)
action \cite{BP},

\[
\left\langle \Phi \right\rangle _{{\rm ex}}=\frac{\Im _{i}}{2}\int \frac{d%
{\bf r}}{2\pi }\frac{|\vec{\partial}w|^{2}+\theta i(\partial _{x}w\partial
_{y}w^{\ast }-\partial _{y}w\partial _{x}w^{\ast })}{\left( 1+|w|^{2}\right)
^{2}}. 
\]
When the latter is taken alone, it is minimized \cite{BP} by the classical
BP skyrmion, $w_{0}=\left( x+i\theta qy\right) /R$, where $R$ is the texture
radius, and for one texture carrying $q=1$, or $q=-1$ real particles, 
\begin{equation}
\left\langle \Phi \right\rangle _{{\rm ex}}=\frac{\Im _{i}}{2}\left[ |q|+q%
\right] .  \label{asymmetry}
\end{equation}

We would like to emphasize that the above calculation results in only one
part of the total thermodynamic potential of the electron liquid, $\Phi
=\left\langle E-\mu N\right\rangle $: namely, the exchange energy of a
texture. Since the total number of particles in the liquid is changed by the
texture, an extra term $-\mu q$ has to be added to $\left\langle \Phi
\right\rangle _{{\rm ex}}$ in order to obtain the actual $\Phi $. Besides
that, the 'excess' charge density, $\rho $ carried by the texture with $R\gg
\lambda $ is formed by many electrons, so that the thermodynamic potential
has to be completed by the direct Coulomb energy term \cite{MacDonald}, 
\[
\Phi _{{\rm C}}=\frac{e^{2}}{2\chi }\int d{\bf r}d{\bf r}^{\prime }\frac{%
\rho ({\bf r})\rho ({\bf r}^{\prime })}{|{\bf r}-{\bf r}^{\prime }|}=\frac{%
\xi e^{2}}{\chi R},\;\;\xi \approx \frac{3\pi ^{2}}{64}.
\]

One also has to bring forward the excess Zeeman energy of a texture, 
\begin{equation}
\Phi _{{\rm Z}}=\frac{\varepsilon _{{\rm Z}}}{2}\int (1-n^{z})\frac{d{\bf r}%
}{2\pi \lambda ^{2}}=\int \frac{d{\bf r}}{2\pi }\frac{|\varepsilon _{{\rm Z}%
}|/\lambda ^{2}}{\left( 1+|w|^{2}\right) }.  \label{Zeeman}
\end{equation}
Since, for the BP skyrmion solution, $w_{0}$, the integral in Eq. (\ref
{Zeeman}) logarithmically diverges at long distances, one has to study a
modification of a topological texture by a conformally non-invariant $\Phi _{%
{\rm Z}}$. To that end, we expand the space of functions $w$ onto $%
w=(x+i\theta qy)/[Rf(r)]$, which incorporates a smooth cut-off function $%
f(r)<1$, $f(0)=1$ and $f(\infty )=0$, and has the same topological charge $%
q=\pm 1$ and nominal radius $R$, as $w_{0}$. We find the cut-off function $%
f(r)$ by minimizing the functional$\ \left\langle \Phi \right\rangle _{{\rm %
ex}}+\Phi _{{\rm Z}}$ with respect to $f(r)$, which results in $%
f(r>R)\approx zK_{1}(z)$, where $z=(r/\lambda )\sqrt{2\varepsilon _{{\rm Z}%
}/\Im _{i}}$ and $K_{1}$ is the Bessel function, and gives us the minimal
value for $\left\langle \Phi \right\rangle _{{\rm ex}}+\Phi _{{\rm Z}}$: 
\[
\frac{\Im _{i}}{2}\left[ |q|+q\right] +\frac{\varepsilon _{{\rm Z}}}{2}%
\left( \frac{R}{\lambda }\right) ^{2}\ln \left( \frac{\eta \Im _{i}}{%
\varepsilon _{{\rm Z}}R^{2}/\lambda ^{2}}\right) , 
\]
where $\eta =2e^{-2C}$, $C$ is the Euler number.

As the last step, we find the radius, $R$ of a texture by optimising the sum 
$\left\langle \Phi \right\rangle _{{\rm ex}}+\Phi _{{\rm Z}}(R)+\Phi _{{\rm C%
}}(R)$ for a given $q$, so that

\[
\frac{R}{\lambda }=\left( \frac{\xi E_{{\rm C}}}{\varepsilon _{{\rm Z}}}%
\right) ^{1/3}\left( \ln \frac{\Im _{i}\eta /\xi ^{2/3}}{E_{{\rm C}%
}^{2/3}\varepsilon _{{\rm Z}}^{1/3}}\right) ^{-1/3},\; 
\]
and we find that for a liquid with $N_{{\rm 1}}$ skyrmions and $N_{{\rm -1}}$
antiskyrmion the thermodynamic potential is equal to 
\begin{equation}
\Phi =\sum_{q=\pm 1}\left( \frac{\Im _{i}}{2}\left[ |q|+q\right] +{\cal E}%
|q|-\mu q\right) N_{q},\;{\cal E}=\frac{3\xi }{2}\frac{e^{2}}{\chi R}.
\label{Phisk}
\end{equation}

Now, we can study the energy, $\left\langle E\right\rangle $ of the QHF with
a small number of charged excitations of all types: $N_{{\rm e}}$
single-particle (un-screened) electrons, $N_{{\rm h}}$ 'plain' holes, $N_{1}$
skyrmions and$\ N_{-1}$ anti-skyrmions (screened holes) added to exactly $%
\nu =2N+1$ state under the electrical neutrality condition, $N_{{\rm e}%
}+N_{1}=N_{{\rm h}}+N_{-1}$. Using expressions in Eqs. (\ref{Eeh}),(\ref
{Phisk}) and the fact that $N=N_{{\rm e}}+N_{1}-N_{{\rm h}}-N_{-1}=0$, we
find that 
\begin{equation}
\left\langle E\right\rangle =\varepsilon _{{\rm Z}}N_{{\rm e}}+{\cal E}%
N_{-1}+(\Im _{i}+{\cal E})N_{1}+\Im _{0}N_{{\rm h}}.  \label{SkEnergy}
\end{equation}

The latter equation shows that when, formally, ${\cal E}<\Im _{0}$ and $R\gg
\lambda $ (which is a stronger restriction), the activation energy of a
neutral pair of charged carriers in the thermodynamic limit is equal to $%
{\cal E}$ and is determined by the thermally activated pair consisting of an
electron and an antiskyrmion. Under the approximation specified above, this
activation energy depends on the inter-LL exchange energy only weakly, that
is, it is almost insensitive to the extent of the electron wave function
across the heterostructure or quantum well, and also to the number $K$ of
the highest occupied Landau level in the QHF ($\nu =2K+1$). The same energy
parameter also determines the jump of the chemical potential of a liquid 
\cite{Dolgopolov} upon the variation of the filling factor across the
integer value. The result specified in Eq. (\ref{Result}) agrees with the
activation transport gaps measured in heterostructures with a vanishing
Zeeman splitting \cite{Leadley,Leadley1,Shayegan}. It also explains a small
value of ${\cal E}$ observed in narrow quantum wells \cite{Maude}, where the
limit $\varepsilon _{{\rm Z}}=0$ is partly hindered by spin-orbit coupling
effects \cite{FI}. However, the energy factors which determine dynamical
properties of QHF's may differ very strongly and require a further analysis.

Authors thank R.Nicholas, A.MacDonald, N.Cooper and I.Kukushkin for
discussions. This work was supported by EPSRC, NATO CLG and EU High Field
ICN.


\begin{references}
\bibitem{RemU2}  We use transformations from the U(2) group to be able to
provide a non-singular mapping of a plane onto the group, so that $%
U(\partial _{x}\partial _{y}-\partial _{y}\partial _{x})U^{\dagger }=0$
within the plane, with all imaginable singularities \cite{Volovik} removed
to the infinity.
\bibitem{Rem1}  To expand, one should use a definite form of the
single-particle Green functions from Eq. (\ref{eq7}). In the imaginary time
representation \cite{Fradkin}, the latter is $G_{0}\left( \tau ;\tau
^{\prime }\geq \tau \right) =-\Lambda e^{-(\tau -\tau ^{\prime })(-\Im
_{0}-\mu )}\sum_{p}|0,p\rangle \langle 0,p|$; $G_{0}\left( \tau ;\tau
^{\prime }<\tau \right) =\Lambda \sum_{n>0,p}e^{-(\tau -\tau ^{\prime
})(n\omega _{c}-\Im _{n}-\mu )}|n,p\rangle \langle n,p|$ $+\left( 1-\Lambda
\right) \sum_{n\geq 0,p}e^{-(\tau -\tau ^{\prime })(n\omega _{c}-\mu
)}|n,p\rangle \langle n,p|,$ where $\Im _{n}$ is the exchange energy between
one electron at the $n$th and the sea of electrons at $0$th LL, and $\mu $
is determined with respect to the position of the lowest LL.
\end{references}
\end{document}